\documentclass[iop,apj]{emulateapj}
\usepackage{epstopdf}

\shortauthors{Ruan, Anderson, MacLeod et al.}
\shorttitle{The Optical Variability of Bright Blazars}
\begin{document}
\title{Characterizing the Optical Variability of Bright Blazars: Variability-Based Selection of \emph{Fermi} AGN}
\author{John~J.~Ruan\altaffilmark{1,2}, 
  Scott~F.~Anderson\altaffilmark{2}, 
  Chelsea~L.~MacLeod\altaffilmark{3,2}, 
  Andrew~C.~Becker\altaffilmark{2},
  T.~H.~Burnett\altaffilmark{4}, 
  James~R.~A.~Davenport\altaffilmark{2},
  \v{Z}eljko~Ivezi\'{c}\altaffilmark{2}, 
  Christopher~S.~Kochanek\altaffilmark{5},
  Richard~M.~Plotkin\altaffilmark{6,7},  
  Branimir~Sesar\altaffilmark{8},
  J.~Scott~Stuart\altaffilmark{9}
  }
\altaffiltext{1}{ Corresponding author: jruan@astro.washington.edu} 
\altaffiltext{2}{Department of Astronomy, University of
Washington, Box 351580, Seattle, WA 98195, USA}
\altaffiltext{3}{Department of Physics, United States Naval Academy, 572C 
Holloway Road, Annapolis, MD 21402, USA }
\altaffiltext{4}{Department of Physics, University of Washington, Seattle, WA 98195-1560, USA}
\altaffiltext{5}{Department of Astronomy, Ohio State University, 140 West 18th Avenue, Columbus, OH 43210, USA} 
\altaffiltext{6}{Department of Astronomy, University of Michigan, 500 Church Street, Ann Arbor, MI 48109, USA} 
\altaffiltext{7}{Astronomical Institute Anton Pannekoek, University of Amsterdam, Science Park 904, 1098 XH, Amsterdam, The Netherlands} 
\altaffiltext{8}{Division of Physics, Mathematics and Astronomy, Caltech, Pasadena, CA 91125, USA} 
\altaffiltext{9}{Lincoln Laboratory, Massachusetts Institute of Technology, 244 Wood Street, Lexington, MA 02420-9108, USA} 


\begin{abstract}
	We investigate the use of optical photometric variability to select and identify 
blazars in large-scale time-domain surveys, in part to aid in the identification of blazar 
counterparts to the $\sim$30\% of $\gamma$-ray sources in the \emph{Fermi} 
2FGL catalog still lacking reliable associations. Using data from the optical LINEAR 
asteroid survey, we characterize the optical variability of blazars by fitting a damped 
random walk model to individual light curves with two main model parameters, the 
characteristic timescales of variability $\tau$, and driving amplitudes on short 
timescales $\hat{\sigma}$. Imposing cuts on minimum $\tau$ and $\hat{\sigma}$ 
allows for blazar selection with high efficiency $E$ and completeness $C$. To test 
the efficacy of this approach, we apply this method to optically variable LINEAR  
objects that fall within the several-arcminute error ellipses of $\gamma$-ray sources 
in the \emph{Fermi} 2FGL catalog. Despite the extreme stellar contamination at the 
shallow depth of the LINEAR survey, we are able to recover previously-associated 
optical counterparts to \emph{Fermi} AGN with $E$ $\ge$ 88\% and $C$ $=$ 88\% 
in \emph{Fermi} 95\% confidence error ellipses having semimajor axis $r$ $<$ 8$\arcmin$. 
We find that the suggested radio counterpart to \emph{Fermi} source 
2FGL J1649.6+5238 has optical variability consistent with other $\gamma$-ray 
blazars, and is likely to be the $\gamma$-ray source. Our results suggest that the 
variability of the non-thermal jet emission in blazars is stochastic in nature, with 
unique variability properties due to the effects of relativistic beaming. After correcting
for beaming, we estimate that the characteristic timescale of blazar variability is 
$\sim$3 years in the rest-frame of the jet, in contrast with the $\sim$320 day disk
flux timescale observed in quasars. The variability-based selection method presented 
will be useful for blazar identification in time-domain optical surveys, and is also a probe of 
jet physics. 
\end{abstract}

\section{Introduction}

	Blazars are a relatively rare sub-class of active galactic nuclei (AGN) in which a 
jet is aligned along the observer's line of sight, leading to the effects of relativistic 
beaming and unusual associated emission \citep{bl78}. Blazars are 
among the most variable extragalactic objects detected in time-domain 
optical surveys, and have strong emission from radio to TeV energies \citep{ul97}. 
The central engine is believed to be accretion onto a super-massive black hole, 
driving relativistic outflows in a collimated jet with typical Lorentz factors on the order
of $\Gamma$ $\sim$ 10 that is pointed to within angle $\Gamma^{-1}$ of the observer. 
The term `blazars' usually encompasses both BL Lac objects and Flat-Spectrum Radio 
Quasars (FSRQs), which are believed to be jet-aligned Faranoff-Riley Type I and 
II AGN, respectively \citep{ur95}. In this paper, we adopt this definition.
	
	The canonical broadband spectral energy distribution (SED) of blazars typically 
includes several main components: (1) a synchrotron peak likely due to tangled magnetic fields 
in the jet that may extend from the radio to the soft X-ray regime; (2) an inverse-Compton 
peak in the X-ray to GeV regime likely due to scattering of synchrotron or external 
photons off of relativistic electrons in the jet; (3) a possible inverse-Compton component 
in the soft X-rays due to a hot corona; and (4) occasional hints of an underlying 
accretion disk continuum or host galaxy emission. The optical and $\gamma$-ray
observations we use in this paper are expected to be dominated by the synchrotron
and jet inverse-Compton emission, respectively.
	
	The strong high-energy inverse-Compton peak in the SEDs of blazars causes them to 
account for the vast majority of bright extragalactic $\gamma$-ray emitting sources. 
The \emph{Fermi Space Telescope} has surveyed the $\gamma$-ray sky since launch in 
2008, and its Large Area Telescope \citep[LAT,][]{at09} instrument provides by far the deepest survey 
to date in the 100 MeV - 100 GeV regime. The current 2-year LAT source catalog 
\citep[2FGL,][]{no12} includes 1873 total sources, $\sim$44\% of which are reliably 
associated with AGN, and an additional $\sim$14\% are candidate 
AGN associations. Of the reliably associated AGN, the overwhelming majority are blazars.
Approximately half of these \emph{Fermi} blazars are BL Lac objects, 
and half are FSRQs \citep{ac11}.
	
	The 2FGL catalog is a significant improvement over the 11-month \emph{Fermi}
LAT source catalog \citep{ab10} in 
number of sources, source detection methods, and source associations to known objects. 
However, $\sim$32\% of sources in the 2FGL catalog still lack 
reliable associations with any known object, many of which may be blazars, especially 
at high Galactic latitudes \citep{ac12}. Although the angular resolution of the LAT is a dramatic 
improvement over previous all-sky $\gamma$-ray survey instruments such as EGRET, typical 95\% 
confidence error ellipses of sources in the 2FGL catalog are still on the order 
of several arcminutes in size and can contain numerous candidate counterparts 
at other wavelengths, making association and identification of $\gamma$-ray sources difficult. 
	
	 Recently, huge efforts have gone into identifying \emph{Fermi} $\gamma$-ray 
sources by positional coincidences of candidate counterparts from multi-wavelength 
surveys \citep[e.g.][]{ste10, mae11}, observations of contemporaneous variability 
at different wavelengths \citep[e.g.][]{ka12}, and statistical methods based on 
observed $\gamma$-ray source properties \citep[e.g.][]{ac12}. Notably, \citet{mas11} 
applied a variant of the well-known method of selecting AGN by their unique colors 
in the mid-IR \citep[e.g.][]{la04, st05} to data from the \emph{WISE} survey 
\citep{wr10}, and used their selection method to separate blazars from stars. 
\citet{mas12} then further utilized their technique to find \emph{WISE}-detected 
$\gamma$-ray blazar candidates in the 2FGL, achieving excellent completeness, 
although the efficiency of their method is unclear (see discussion in Section 5.1).

	Aside from their distinct mid-IR colors, blazars are also unique in their 
strong variability at nearly all wavelength regimes. This will make blazars stand 
out in the flood of time-domain data from current and future large-scale optical 
imaging surveys such as Pan-STARRS \citep{ka10}, the Catalina 
Real-Time Transient Survey \citep{dr09}, the Palomar Transient  Factory 
\citep[PTF,][]{ra09}, and the Large Synoptic Survey Telescope \citep[LSST,][]{iv08}. 
It is thus an auspicious time to explore the possibility of selecting and identifying 
$\gamma$-ray emitting blazars \emph{en masse} in \emph{Fermi} error ellipses by 
their optical photometric variability. 

	Blazars have observed variability on timescales from hours to 
years, and their distinctiveness in time-domain surveys results from two effects. 
Firstly, the relativistic jet strongly beams non-thermal emission along the line of 
sight towards the observer, boosting the luminosity by several orders of 
magnitude. Secondly, relativistic Doppler-boosting shortens the observed 
timescale of variability in comparison to that in the rest-frame of the 
outflow. However, the physical mechanisms causing the variability are far less 
clear. Since blazar emission is largely dominated by non-thermal emission from 
the jet, it is possible that internal shocks from overtaking collisions of fluid shells 
with different velocities along the jet produce time-variable synchrotron 
emission \citep[e.g.][]{bo10}, although other emission components such as the 
accretion disk may also be time-variable. 

	Other classes of AGN, including the far more numerous `normal' quasars
(i.e. non-FSRQ, type 1 AGNs), are also highly optically variable, 
although the predominance of particular physical variability mechanisms may differ among 
AGN subclasses. \citet{se07} have shown that $\sim$90\% of quasars in Sloan 
Digital Sky Survey \citep[SDSS,][]{yo00} Stripe 82 are variable at the $>$0.03 intrinsic 
rms variability level (defined in Eq. 1), and previous studies
using large samples of individual quasar light curves have shown that quasar 
optical variability is well described by a damped random walk (DRW) model 
\citep{ke09, ma10, ko10, bu11, zu12}. \citet[][hereafter MA11]{ma11} further utilized the 
DRW model as a tool to separate stars and quasars by optical photometric variability 
in SDSS Stripe 82, and showed that quasar selection using this method can 
achieve efficiency $E$ $\geq$ 80\% and completeness $C$ $=$ 90\% by 
variability alone. This is particularly useful for selection of quasars in certain 
redshift regimes, where color selection may fail due to overwhelming contamination 
from foreground stars. Furthermore, studies of the long-term variability characteristics of 
AGN may also provide a unique perspective on accretion disk and jet physics.

	The optical variability of small numbers of individual blazars has been 
studied extensively in the literature through intensive long-term monitoring campaigns 
\citep[e.g.][]{we88, ca92}. However, no consistent picture of variability has emerged, 
at least partially due to the heterogenous nature of these observations and 
the small sample sizes. Blazar light curves from a large-scale flux-limited time-domain 
survey should be more homogenous, and the 
depth of current surveys can yield orders of magnitude more blazar light curves 
than possible through targeted monitoring campaigns. However, 
the rarity of blazars places significant constraints on the specifications of 
any time-domain survey in which a large number of blazar light curves can be obtained. 
Despite the success of the SDSS Stripe 82 for time-domain astronomy, the depth of 
the survey ($\sim$22.5 magnitude in $r$ band) does not compensate for the 
small sky coverage ($\sim$290 deg$^2$) when it comes to studying blazars.
We instead employ the recalibrated LINEAR survey, which essentially covers the 
$\sim$10,000 deg$^2$ SDSS photometric footprint down to $\sim$17 magnitude 
in $r$ band \citep[][hereafter SE11]{se11}. \citet{ba09a} provided a previous study of blazar optical 
variability in the Palomar-Quest Survey, and \citet{ba09b} conducted a subsequent 
variability-based search for new blazars within that survey. However, the 
Palomar-Quest survey provided a median of $\sim$5
epochs of observation for each object per filter, and could only be used to study the ensemble 
variability of blazars rather than the individual sources. By contrast, the $\sim$200 
epochs of observation per object in the LINEAR survey allow us to directly model the 
light curves of individual blazars in this study.

	The structure of this paper is as follows: In Section 2, we describe the LINEAR 
survey and the construction of the Variable LINEAR catalog, from which our blazar and 
comparison normal quasar samples are drawn. In Section 3, we describe the damped 
random walk model of variability, and our results of modeling the LINEAR 
light curves of blazars, normal quasars, and stars. In Section 4, we test the efficiency and 
completeness of blazar selection using a DRW model within \emph{Fermi} error ellipses. 
In Section 5, we discuss this method in the context of its future 
applications, as well as possible implications of our work for AGN physics. 
We summarize and conclude in Section 6.

\section{The LINEAR Survey}

\subsection{Photometric Data}	

	All photometric data used in this paper are from the archives of the MIT Lincoln 
Laboratory (MITLL) Lincoln Near-Earth Asteroid Research (LINEAR) survey, spanning 
the period from December 2002 through March 2008. A review of the original LINEAR 
near-Earth asteroid survey program is presented in \citet{st00}, and the subsequent 
photometric recalibration of the archived data using SDSS to construct the LINEAR photometric 
database is discussed in SE11. We summarize only the most 
salient points here. The LINEAR survey program used two 1.01m diameter telescopes 
at the Experimental Test Site within the US Army White Sands Missile Range in New 
Mexico, each equipped with a 5 megapixel (2560$\times$1960) back-illuminated CCD 
developed at MITLL and described in \citet{bu98}. The recalibrated survey 
covers $\sim$10,000 deg$^2$ of sky, overlapping the SDSS photometric footprint,
and contains over 5 billion photometric measurements of $\sim$25 million objects 
down to a 5$\sigma$ depth of $r$ $\lesssim$ 18 mag. Due to the astrometric goals of the 
original survey and to increase S/N, only a single 
broad filter was used. The cadence range spans minutes to years, with a peak at 
the main 15-minute cadence, a gap at 8-hours, and a secondary peak around 11 days. 
The median number of good observations per object in the full LINEAR catalog is 
$\sim$460 within $\pm$10$^\circ$ of the ecliptic plane and $\sim$200 elsewhere. 

	SE11 extracted sources from LINEAR imaging using fixed-aperture photometry, 
and recalibrated the astrometry to the USNO-B catalog \citep{ba08}. To recalibrate 
the photometry using SDSS, SE11 first matched LINEAR sources to SDSS DR7 non-saturated, 
primary objects. Using $gri$ photometry of 
DR7 objects matched to LINEAR, SE11 then modeled and calculated synthetic LINEAR 
magnitudes of matched SDSS objects m$_{SDSS}$, effectively turning the SDSS 
imaging catalog into a catalog of LINEAR photometric standard stars. After a super 
flat-field correction on each field using these calibration stars, a catalog of 5 billion 
individual point sources with recalibrated LINEAR magnitudes m$_{LINEAR}$ from all 
epochs of observations was positionally clustered into 25 million objects, each with at 
least 15 epochs of observation. This comprises the full recalibrated LINEAR catalog. Checks on 
the recalibrated photometry by SE11 show that the median m$_{LINEAR}$ $-$ m$_{SDSS}$ 
residual per field has a distribution of about 0.01 mag wide. This should should not be 
confused with the single-epoch photometric uncertainties for individual objects in 
LINEAR, which are generally $\gtrsim$0.04 mag (see Figure~\ref{fig:photoerrors}). Fields with 
m$_{LINEAR}$ $-$ m$_{SDSS}$ $>$   0.1 (usually due to variable cloud coverage) 
are removed.	
	
\subsection{The Bright LINEAR Catalog}

	From the full LINEAR catalog, we select a bright subset catalog suitable for variability
science by imposing a cut on the minimum number of good observations in each light 
curve of $\ge$30, and a LINEAR recalibrated magnitude cut of 14 $<$ $m_{LINEAR}$ $<$ 17 
(where photometric errors are $\lesssim$ 0.11 mag).
At magnitudes $>$17 and $<$14, single epoch photometric errors rise rapidly due to photon noise 
and saturation, respectively. These cuts provide us with  $\sim$4.5 million objects in 
the Bright LINEAR catalog. 
	
	To create a sample of known blazars and quasars in the Bright LINEAR catalog, we 
positionally match the LINEAR sources to catalogs of known BL Lac objects and 
FSRQs. To find the known BL Lac objects, we match 
to 1,371 BL Lac objects in the \citet{ve10} catalog, 501 BL Lac objects in the 
\citet{pl08} catalog, and 637 radio-loud BL Lac objects in the \citet{pl10} catalog (none
of the 86 radio-quiet objects in \citet{pl10} matched to a LINEAR object), all using a 
$3\farcs0$ matching radius, resulting in 101 matched distinct BL Lac objects. Known 
quasars in the Bright LINEAR catalog were identified by matching to the 105,782 
quasars in the \citet{sc10} SDSS Data Release 7 catalog, also with a $3\farcs0$ matching 
radius, resulting in 1,020 matched distinct quasars. We also separate out 
the FSRQs by positionally matching these 1,020 quasars to the CRATES radio survey 
\citep{he07} of 11,131 flat-spectrum ($\alpha > -0.5$, where $S_\nu \propto \nu^\alpha$) 
radio sources using a $3\farcs0$ matching radius, resulting in 42 distinct FSRQs. The 978
non-flat radio spectrum quasars remaining are classified here as normal quasars (i.e. 
unlikely to be strongly dominated in the optical by jet emission). 
	
	We note that there is ultimately an overlap of 3 objects between the BL Lac object
and FSRQ samples due to double counting objects which appear in the BL Lac object
catalogs, the \citet{sc10} quasar catalog, as well as the \citet{he07} catalog, which may 
result from the intermittent appearance of broad emission lines in their optical spectra. 
Since our definition of `blazars' here includes both BL Lac objects and FSRQs,
we count these 3 objects as blazars (but not normal quasars due 
to their flat radio spectra). In summary, we identified a total of 140 blazars and 
978 normal quasars in the Bright LINEAR catalog.
	
\begin{figure}
\centering
\includegraphics*[width=0.47\textwidth]{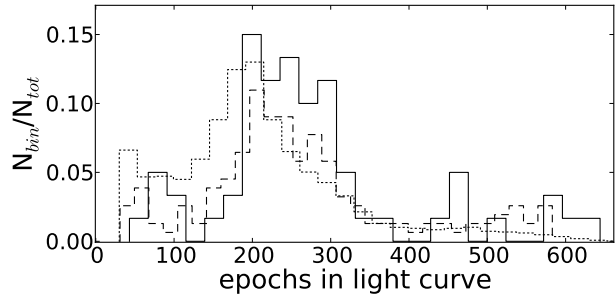}
\caption{
Normalized histograms of the number of light curve
epochs available for each source in the Bright LINEAR catalog. The distributions are
shown for 140 blazars (solid), 978 normal quasar (dashed), and all other
objects (dotted).
}
\label{fig:nptgood}
\end{figure}

	The potential of the LINEAR survey for blazar time-domain studies can be seen in
Figure~\ref{fig:nptgood}, which shows normalized histograms of the number of LINEAR 
epochs of observation for each of the 140 objects in the blazar sample, the 987 objects 
in the normal quasar sample, and all other objects (mainly stars) in the Bright LINEAR Catalog.
The histograms are all generally consistent with each other, peak at $\sim$200 epochs, and 
lack objects with $<$30 good observations due to the previously imposed cut. 

\begin{figure}
\centering
\includegraphics[width=0.47\textwidth]{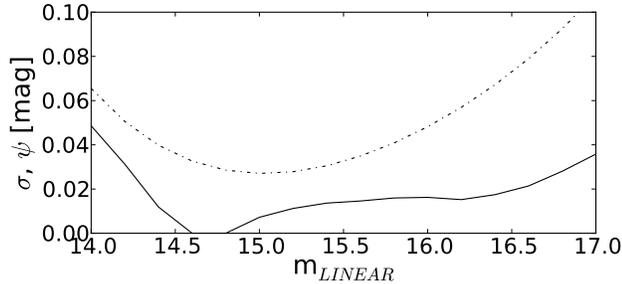}
\caption{
Median intrinsic rms variability $\sigma$ (solid) and photometric uncertainty $\psi$(m$_{LINEAR}$) (dash-dot; Equation 7 from SE11) of LINEAR objects in the Bright LINEAR catalog, as a function of recalibrated LINEAR magnitude.
}
\label{fig:photoerrors}
\end{figure}

	Before  attempting to model the light 
curve of every object, we would first like to compare the general level of 
variability of blazars, normal quasars, and other objects in the Bright
LINEAR Catalog. We follow \citet{se07} and estimate the intrinsic rms variability 
$\sigma$ for each light curve, defined as
\begin{equation}
	\sigma = [\Sigma^2 - \psi(m_{LINEAR})^2]^{1/2}
\label{eqn:sigma}
\end{equation}
for all objects with $\Sigma$ $>$  $\psi$(m$_{LINEAR}$)  and $\sigma$ = 0 otherwise, where 
$\Sigma$ is the rms scatter of the m$_{LINEAR}$ for all observations of 
each object, and $\psi$(m$_{LINEAR}$) is the  median photometric error of 
LINEAR objects as a function of magnitude found by SE11 (see their Equation 7). 
Since the photometric errors
in LINEAR are relatively large, $\Sigma$ mainly reflects photometric noise rather
than variability for the vast majority of objects. Equation 1 attempts to remove the 
effects of the median photometric error, and thus $\sigma$ is a better measure of 
intrinsic variability. Figure~\ref{fig:photoerrors} shows the median $\sigma$ and 
$\psi$(m$_{LINEAR}$ ) of all objects in the Bright LINEAR 
catalog as a function of m$_{LINEAR}$ magnitude. 

\begin{figure}
\centering
\includegraphics*[width=0.47\textwidth]{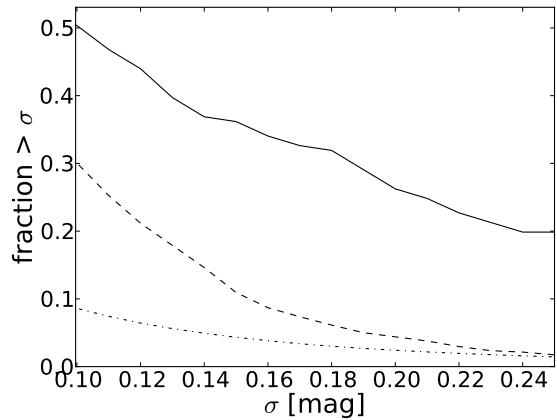}
\caption{ 
Fraction of objects with a intrinsic rms variability larger than $\sigma$, as a function of $\sigma$, for blazars (solid), normal quasars (dashed), and other objects (dash-dot). 
}
\label{fig:rmsfract}
\end{figure}

	We compare the variability of blazars to normal quasars and other 
LINEAR objects, as it is expected that blazars should be systematically 
more variable. Figure~\ref{fig:rmsfract} shows the integrated distribution of blazars,
normal quasars, and other LINEAR bright objects in the Bright LINEAR catalog
as a function of the estimated intrinsic rms variability $\sigma$. We estimate that 
$\sim$36\% of blazars, as compared to only $\sim$11\% of quasars, are 
variable with $\sigma$ $>$ 0.15 mag in this passband. At this level of variability, 
the estimate of $\sigma$ is not significantly affected by uncertainties in 
$\psi$(m$_{LINEAR})$. Only $\sim$4\% of the other bright 
LINEAR objects (non-blazars and non-quasars) are variable above this 
 $\sigma$ $>$ 0.15 mag level, and they are likely a combination of
 variable stars (e.g. RR Lyrae), eclipsing binaries, underestimates 
of the photometric errors $\psi$, source blending, or other photometric issues. 

\subsection{The Variable LINEAR catalog}

	In order to characterize the photometric variability, we need a clean sample of 
light curves dominated by intrinsic variability rather than photometric errors. 
Thus, we follow the method of \citet{se07} to create a clean Variable LINEAR catalog 
as a subsample of the 4.5 million objects in the Bright LINEAR catalog discussed 
in the previous section.

	As a first variability criterion, we require $\sigma >$ 0.1, roughly 
equivalent to selecting sources with $\sigma$ more than two times the 
measurement noise at 14 $<$ m$_{LINEAR}$ $<$ 16, and roughly corresponding to the 
photometric error at m$_{LINEAR}\sim17$. Approximately 8\% of objects in the Bright 
LINEAR Catalog pass this initial selection cut, including some non-variable objects at the 
faint end with large $\sigma$ due to large photometric errors. To reduce these 
spurious contaminants, we assume that the photometric error distribution follows 
a Gaussian, and place a cut on the $\chi^2$ per degree of freedom calculated with respect 
to a weighted mean magnitude and errors from the photometry. A cut of $\chi^{2}/dof$ $>$ 3 
leaves 188,745 objects comprising the Variable LINEAR catalog we use hereafter, each 
typically having $\sim$200 epochs per light curve.

	Matching this variable LINEAR sample to the various catalogs described in Section 
2.2 yields 60 blazars and 155 normal quasars, including 1 of the 3 overlapping objects 
discussed previously in Section 2.2 that we include in both
samples. Although the variable LINEAR sample cuts out the majority of the 
known 140 blazars and 978 normal quasars in the Bright LINEAR catalog,  
this is not because the majority of blazars and normal quasars are non-variable, but 
rather because the large photometric errors in LINEAR overwhelm the intrinsic variability 
of fainter LINEAR objects. This effect can be seen in the much larger 
fraction of blazars (60 out of 140) that survive these variability cuts, in comparison to normal 
quasars (155 out of 978). Since blazars tend to be systematically more variable than normal 
quasars (as we have shown in Section 2.2), their intrinsic variability will dominate over 
photometric errors for a much larger fraction of objects.  

\section{The Damped Random Walk Model}

	MA11 designed and tested optical variability-based selection of quasars in SDSS 
Stripe 82 by modeling the individual light curves of known quasars as a damped random 
walk. We perform a similar analysis of the LINEAR light curves of known blazars and normal 
quasars, and show that blazar light curves lie in a distinct region of DRW variability parameter 
space. The DRW model statistically parametrizes variability using three parameters: a mean 
light curve magnitude, a characteristic damping timescale of variability $\tau$, and a driving amplitude 
of short-term stochastic variability $\hat{\sigma}$ \citep[see][for further discussion]
{ke09, ko10, ma10, zu12}. The fitted parameters can be expressed in terms of the structure 
function (SF), defined as the rms magnitude difference between all pairs of observations in 
each individual light curve as a function of the time-lag. While this model is phenomenological, 
it is likely that the fitted parameters reflect the physical processes that cause the variability. 

	Following \citet{ko10} and MA11, we model each individual light curve in our sample 
as a stochastic process described by the exponential covariance matrix
\begin{equation}
S_{ij} = \sigma^{2}\mathrm{exp}(-|t_{i}-t_{j}|/\tau)
\label{eqn:covmatrix}
\end{equation}
for each pair of observations at time $t_i$ and $t_j$ in the light curve. This describes a 
DRW process with a characteristic damping timescale $\tau$ beyond which the structure 
function will asymptote to a constant value of SF$_\infty = \hat{\sigma}\sqrt{\tau} $, and 
a long-term standard deviation of variability $\sigma = \sqrt{2}$SF$_{\infty}$. The short-term 
driving amplitude of variability $\hat{\sigma} = \sigma \sqrt{2/\tau}$ determines
the rise of SF($\Delta\mathit{t}$) for $\Delta$$t \ll \tau$. We model 
each individual light curve and calculate $\hat{\sigma}$ and $\tau$,
along with their likelihood distributions, using the method of \citet{pr92}, its 
generalization in \citet{ry92}, and the fast computational implementation in \citet{ry95}.	
The corresponding structure function for the model for time-lag $\Delta t$ is
\begin{equation}
 \rm{SF}(\Delta \mathit{t})= \rm{SF}_{\infty}(1-e^{-|\Delta \mathit{t}|/\tau})^{1/2}.
\label{eq:sfdt}
\end{equation}

	In fitting each individual light curve with the DRW model, we calculate the 
likelihood of a DRW solution $L_{DRW}$, as well as the likelihood of a pure white 
noise solution $L_{noise}$. Light curves that are better described by the DRW 
model than white noise at a 5-sigma level will have 
$\Delta$$L_{noise}$ = ln($L_{DRW}$/$L_{noise}$) $>$ 5. All our FSRQ and BL Lac object 
light curves in the Variable LINEAR catalog fit the DRW model at above this 5-sigma level.
To remove light curves for which the survey length is shorter than $\tau$ (thus leaving $\tau$ 
unconstrained), we also calculate the likelihood of a runaway timescale, 
$L_\infty$. Objects for which $\tau$ is constrained in the DRW model will have 
$\Delta$$L_\infty$ $=$ ln($L_{DRW}$/$L_\infty$) $>$ 0.05 \citep{ma10}. \citet{ma10} showed 
that the DRW is an excellent fit to quasar light curves, 
and used cuts on minimum $\Delta$$L_{noise}$ and minimum 
$\Delta$$L_\infty$ to select quasars with high quality light curves. In the higher 
stellar density regions of Stripe 82, MC11 used cuts on maximum $\hat{\sigma}$, 
minimum $\tau$, and minimum $\Delta$$L_{noise}$ to achieve impressive 
completeness $C$ $=$ 93\% (defined as percentage of the total confirmed 
quasars in the sample which are selected) and efficiency $E$ $>$ 78\% (percentage of
quasar candidates selected which are confirmed quasars) in quasar selection. 
Furthermore, \citet{ko10} showed that the method works even in the high
stellar density regions of the Magellanic Cloud fields

	We calculate the best-fit DRW variability parameters of the individual 
light curves of the 60 blazars, 155 normal quasars, and a random sample of 
6000 other objects (mainly foreground Galactic stars, representative of the 
other $>$180,000 objects) from the Variable LINEAR catalog. 
To first understand the underlying distribution of DRW variability parameters for these 
different populations, we redshift-correct our parameters as outlined in \citet{ke09} 
using spectroscopic redshifts from the \citet{sc10} catalog for all FSRQs
and normal quasars. For BL Lac objects, redshifts are much less accurate (and sometimes 
not possible) due to the lack of strong spectral features. Since our sample of BL Lac 
objects is drawn from a variety of catalogs, there are sometimes discordant redshifts 
reported; in such cases, we preferentially adopt the spectroscopically 
derived redshifts (or the lower limits) from the \citet{pl08} and \citet{pl10} catalogs. 
For 3 BL Lac objects, there is no redshift reported in any of the catalogs, and so for these 3 
objects we adopt the mean redshift found for all other LINEAR BL Lac objects, $z$ = 0.28. 
While this assumption is not ideal, very few extragalactic objects in the LINEAR 
survey will be at high redshifts due to the shallow optical flux limit, so uncertainties on these 
3 redshifts will not strongly impact our results.

\begin{figure*}
\centering
\includegraphics[width=0.7\textwidth]{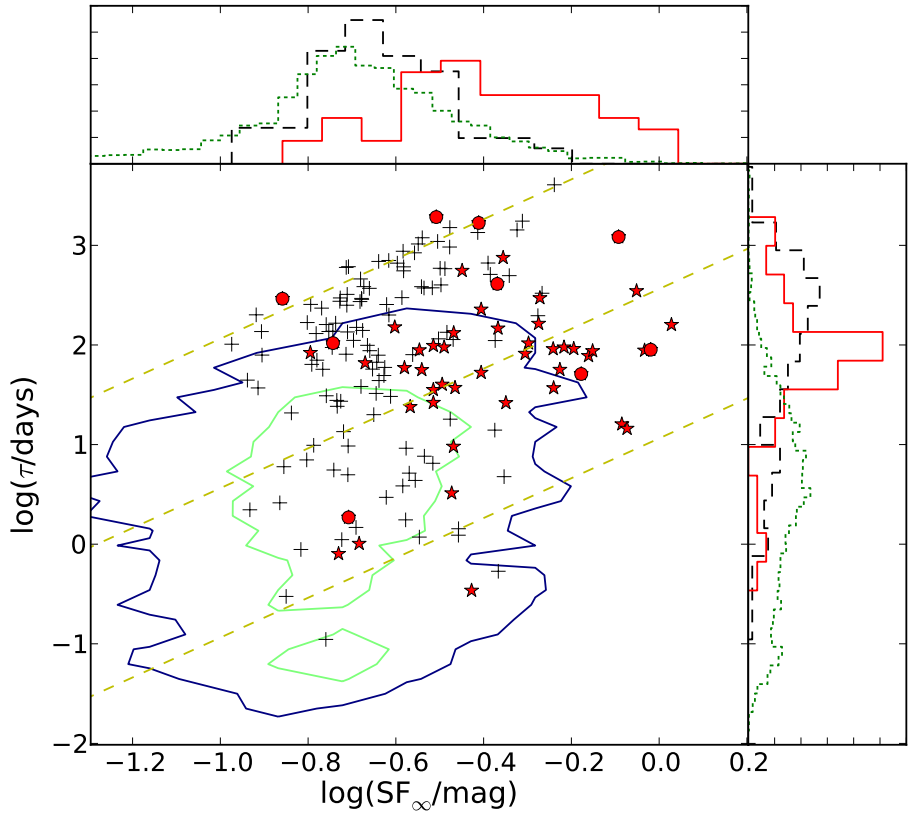}
\caption{Distribution of variable LINEAR objects in SF$_\infty$ (asymptotic value of the structure function on long timescales) and $\tau$ (characteristic damping timescale). BL Lac objects (red stars), FSRQs (red circles), normal quasars (black crosses), and contours enclosing 90\% (blue) and 50\% (green) of all other objects are shown. The side panels show the projected normalized histograms of the distributions in SF$_\infty$ (top) and $\tau$ (right) for blazars (BL Lac objects and FSRQs, red solid), normal quasars (black dashed), and other objects (green dotted). Dashed yellow lines of constant $\hat{\sigma}$ for $\hat{\sigma}$ = $-$0.75, 0.0, 0.75 are also shown, with $\hat{\sigma}$ increasing towards the lower right. 
}
\label{fig:sfinftauvar}
\end{figure*}

\begin{figure}
\centering
\includegraphics[width=0.47\textwidth]{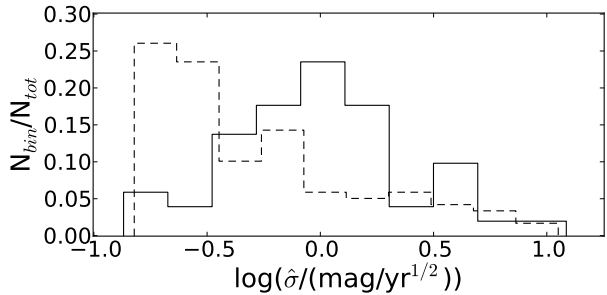}
\caption{
Distributions of the driving amplitude on short timescales $\hat{\sigma}$ for blazars (solid), and normal quasars (dashed) in the Variable LINEAR catalog.
}
\label{fig:sigmahist}
\end{figure}

	Imposing the restrictions on $\Delta L_\infty > 0.05$ and $\Delta L_{noise} > 5$ leaves us with 
51 blazars and 121 normal quasars in our sample. Figure~\ref{fig:sfinftauvar} shows
the best-fit DRW parameters  $\tau$ against SF$_\infty$ of blazars, normal quasars,
and other objects (mostly Galactic stars), along with normalized histograms of their distributions.
Although we use separate symbols for BL Lac objects and FSRQs in Figure~\ref{fig:sfinftauvar},
we do not separate these two blazar sub-populations in our subsequent analysis due to
the small sample size and unknown biases in our blazar sample stemming from possible 
correlations between variability and physical properties of the blazars. Future surveys yielding 
larger samples may be able to robustly constrain differences between these two sub-populations, 
and provide further insight into their jet properties. Figure~\ref{fig:sfinftauvar} is directly 
comparable to the analysis of \citet{ko10} and MA11, and confirms that the structure functions 
of normal quasar light curves tend to have longer 
characteristic timescales of variability $\tau$ and slightly larger SF$_\infty$ than stars. 
More importantly, our analysis also shows that blazars have characteristic timescales 
$\tau$ in between those of normal quasars and other objects (mainly variable stars), as well 
as larger values of SF$_\infty$. The driving amplitude on short timescales $\hat{\sigma}$, 
is related to these two parameters by the relation $\hat{\sigma}$ = 
SF$_\infty$/$\sqrt{\tau}$, as shown in Figure~\ref{fig:sfinftauvar}. 
Figure~\ref{fig:sigmahist} shows the distribution for blazars and normal quasars in
$\hat{\sigma}$, where normal quasars tend to peak at 
log ($\hat{\sigma}$/mag yr$^{-1/2}$)  $\sim -0.75$, 
while blazars tend to peak at log $\hat{\sigma}$ $\sim 0.0$. 
A two-dimensional Kolmogorov-Smirnoff test between the distributions of $\hat{\sigma}$
for blazars and normal quasars in Figure~\ref{fig:sigmahist} gives a p-value
of $2.04\times10^{-5}$. For the distributions of $\tau$ in Figure~\ref{fig:sfinftauvar}, 
the p-value is 0.032.

While MA11 performed quasar selection using this approach by placing cuts on 
maximum $\hat{\sigma}$ and minimum $\tau$, 
we can also see from Figure~\ref{fig:sfinftauvar} that a minimum $\tau$ cut would 
separate blazars from other objects (mostly Galactic stars), while a minimum 
$\hat{\sigma}$ cut can help separate blazars from quasars. However, it is clear 
from Figure~\ref{fig:sfinftauvar} that highly efficient and complete quasar and 
blazar selection purely by optical variability in the LINEAR survey is difficult, 
as there will be either be heavy contamination or a low recovery rate.
The main cause of this is the bright magnitude range (14 $<$ m$_{LINEAR}$ $<$ 17) 
of LINEAR, which overwhelmingly probes Galactic stars and very few actual AGN 
(e.g. there are $>$180,000 sources in the Variable LINEAR catalog, 
only 214 of which are known normal quasars and blazars), and so deeper survey 
data containing a substantial fraction of extragalactic sources is necessary. However, 
currently available data from deeper time-domain optical surveys such as SDSS 
Stripe 82 have comparatively small sky coverage, and do not yield a large 
enough sample of the much rarer blazars.  
	
\section{Blazar Selection in Fermi Error Circles}

	Despite the overwhelming stellar contamination in the bright magnitude range of the 
LINEAR survey, our proposed blazar selection method is still useful with LINEAR
if we restrict our search to the small regions of sky associated with the positions 
of \emph{Fermi} sources. This greatly reduces the stellar contamination, while 
simultaneously providing a test of a novel method (based on long-term 
photometric variability characteristics) to associate $\gamma$-ray sources with 
counterparts at optical wavelengths. We will show the DRW parameters of known 
\emph{Fermi} AGN (overwhelmingly blazars), quantify the completeness and 
efficiency of blazar selection using this method, and identify new variability-selected 
candidate \emph{Fermi} blazar counterparts.

	We first positionally match the Variable LINEAR catalog 
described in Section 2.3 to the 1873 sources in the 2FGL catalog, 
approximating each error ellipse as a circle with radius equal to the semi-major axis of the 
95\% confidence ellipse. This initial match yields 480 variable LINEAR objects in 173 \emph{Fermi} 
error circles. These 480 variable LINEAR optical objects in \emph{Fermi} error circles are then 
positionally matched using a $3\farcs0$ matching radius to the 929 associated AGN in the 2nd Fermi AGN catalog \citep{ac11}, restricting consideration to those that have Bayesian association 
probabilities $>$0.8 \citep[this association method is described in the Appendix of][]{ab10}.
This yields a sample of 47 known (i.e. confidently associated) \emph{Fermi} 
$\gamma$-ray emitting AGN.  We calculate DRW parameters for the LINEAR light curves 
of all 480 variable objects, and make nominals cut on $\Delta L_{noise}$ $>$ 5 and 
$\Delta$$L_\infty$ $>$ 0.05 as in Section 3. 

\begin{figure}
\centering
\includegraphics[width=0.47\textwidth]{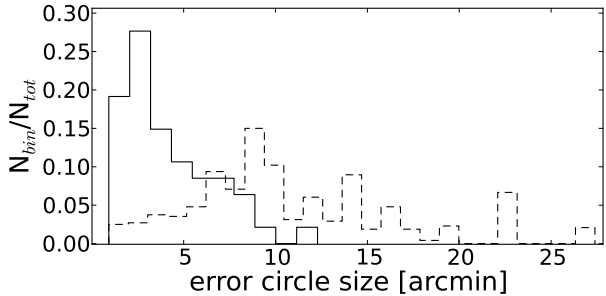}
\caption{Distribution of the number of variable LINEAR objects lying in all \emph{Fermi} error circles of radius $r$ (dashed), and only those lying in \emph{Fermi} error circles of radius $r$ which contain a variable LINEAR object matched to an AGN in the 2nd  \emph{Fermi} AGN catalog (solid), as a function of $r$. 
}
\label{fig:errcirccut}
\end{figure}

	Due to the bright magnitude range of the LINEAR survey, 
it is likely that the flux limit of the LINEAR survey is too bright to actually detect
the optical counterparts of the faintest \emph{Fermi} $\gamma$-ray sources, leaving 
us a population of orphan error circles which contain only LINEAR 
contaminants. Since these fainter sources will also preferentially have larger error circles, 
we can remove them by considering further cuts on the maximum radii of the 
error circles. In Figure~\ref{fig:errcirccut}, we show the distribution of the
total number of LINEAR variable objects in \emph{Fermi} error circles as a function of 
error circle radius $r$, for all 173 error circles that contain at least 1 LINEAR variable 
object, as well as the distribution for the 47 error circles containing a LINEAR variable 
object that matched to an AGN in the 2nd \emph{Fermi} AGN catalog. 
Figure~\ref{fig:errcirccut} shows that the number of variable LINEAR 
objects in error circles with known $\gamma$-ray AGN counterparts peaks at $2\arcmin$
and drops rapidly; beyond $10\arcmin$, there is only 1 \emph{Fermi} error circle that 
contains a known $\gamma$-ray AGN counterpart. However, the distribution of the 
total number of variable LINEAR objects in all error circles as a function of radius has 
a long tail to large $r$, as there are a small number of large error circles ($ r > 15\arcmin$) 
with large numbers of variable LINEAR objects, as expected. This suggests that a cut 
on the error circle radius of $\sim$$10\arcmin$ is appropriate, but to gauge the impact of 
such a cut, we will in detail consider a range of $8\arcmin$, $10\arcmin$ and $12\arcmin$ 
cuts on the maximum error circle size. A cut at $r < 10\arcmin$ removes from consideration 
only 1 \emph{Fermi} error circle containing a known $\gamma$-ray AGN, while a cut at 
$r < 8\arcmin$ removes 4. Scatter in the negative relation between the optical flux and 
$\gamma$-ray error circle size of \emph{Fermi} blazars may introduce some biases in our 
resulting sample towards blazars with higher than average optical to $\gamma$-ray flux ratios when 
we place a cut on the error circle size. However, this bias is expected to be small, as only 
$\sim$7\% of reliably associated AGN in the 2nd Fermi AGN Catalog with optical magnitudes 
$<$17 reside in Fermi error circles $>$10$\arcmin$ in radius. We note that although the 
\emph{Fermi} AGN association success rate of \citet{ac11} as a function of $r$ will also affect 
Figure~\ref{fig:errcirccut}, this effect is secondary to the LINEAR flux limit for the high association 
probabilities considered here. 

\begin{figure*}
\centering
\includegraphics[width=0.47\textwidth]{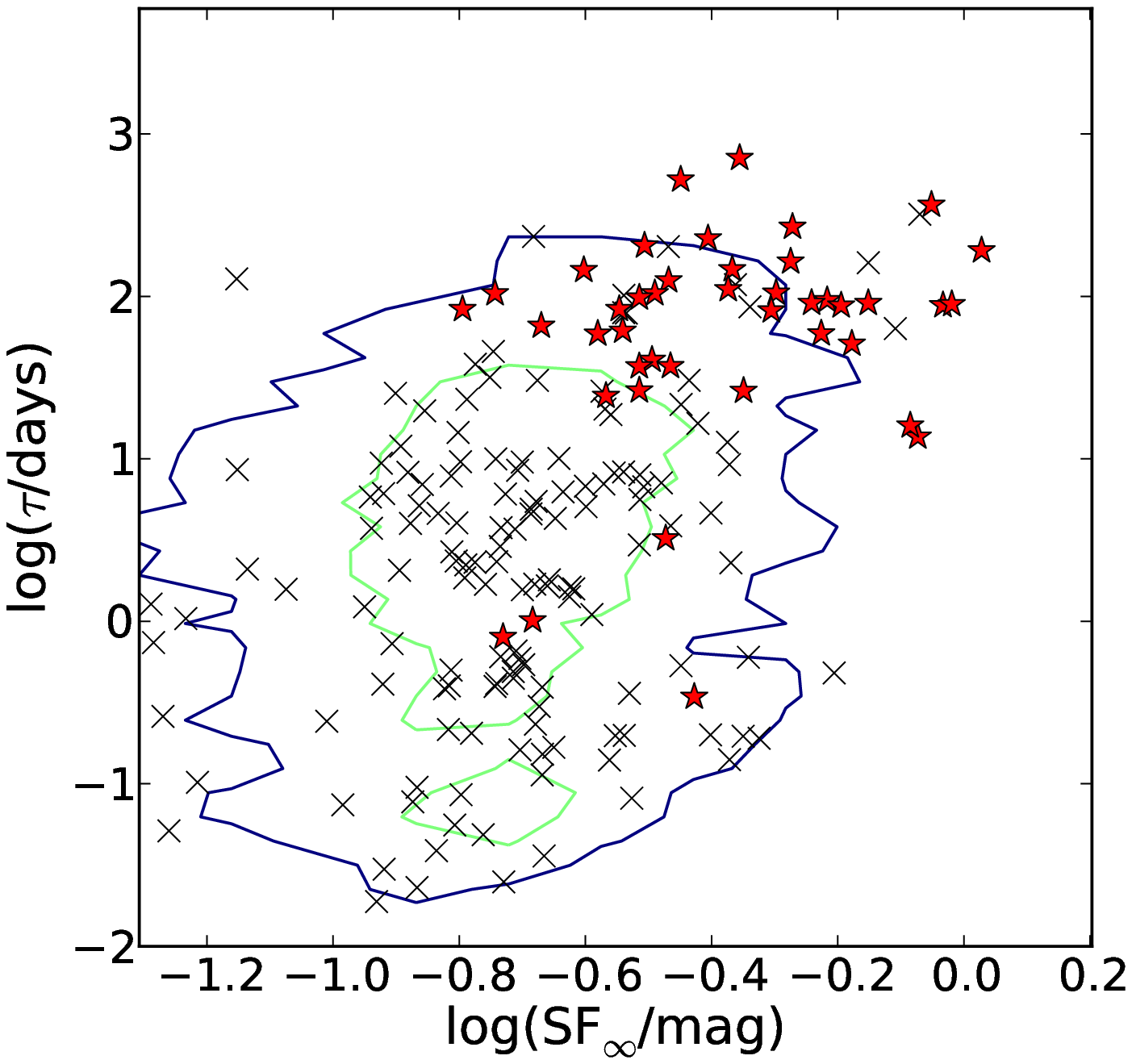}
\includegraphics[width=0.47\textwidth]{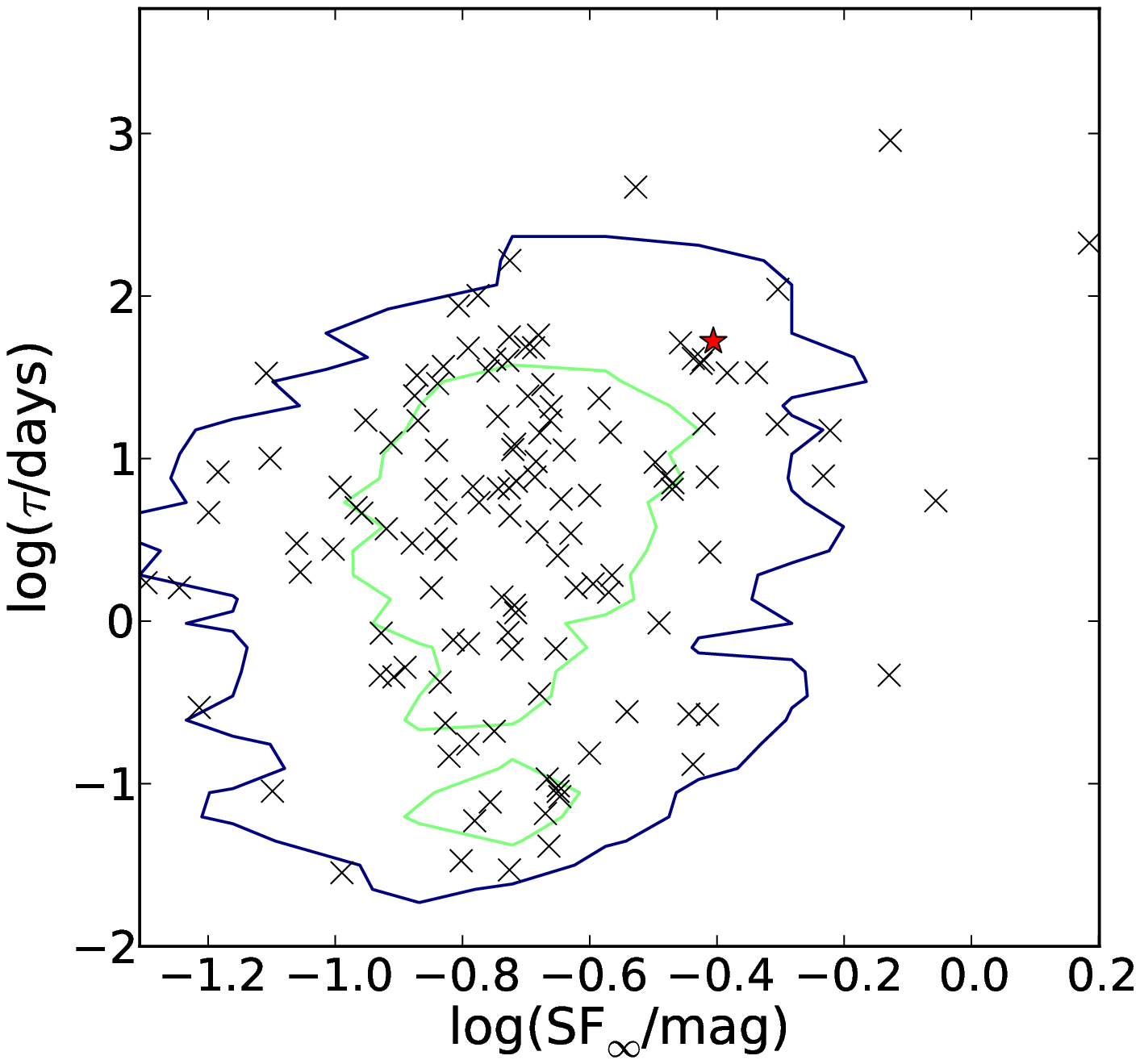}
\caption{ 
\emph{Left}: Distribution of variable LINEAR objects in \emph{Fermi} error circles of radius $r < 10\arcmin$, in SF$_\infty$ and $\tau$, for reliably associated \emph{Fermi} AGN (red stars), and other objects (black crosses). The underlying contours from Figure~\ref{fig:sfinftauvar} enclosing 90\% (blue) and 50\% (green) of all other objects in the
Variable LINEAR catalog is shown in the background for reference. \emph{Right}: Same as the left panel, but for \emph{Fermi} error circles $>$$10\arcmin$ in radius.}
\label{fig:sfinftaufermi}
\end{figure*}

	Figure~\ref{fig:sfinftaufermi} (left panel) shows the distribution in $\tau$ and SF$_\infty$
of LINEAR optical counterparts to \emph{Fermi} $\gamma$-ray AGN, as well as all 
other variable objects in \emph{Fermi} 2FGL error circles of $r < 10\arcmin$. We do 
not apply cosmological redshift-corrections here for blazar selection.
Similar to the conclusions drawn from Figure~\ref{fig:sfinftauvar}, there is clear separation
between $\gamma$-ray emitting AGN and other LINEAR variables in DRW parameter space.
The significant normal quasar population seen in Figure~\ref{fig:sfinftauvar} is 
minimized when looking only in \emph{Fermi} error circles. As expected, almost all 
$\gamma$-ray AGN recovered in our sample have blazar-like variability. 
Figure~\ref{fig:sfinftaufermi}  (right panel) shows that variable objects in error circles with radius $r > 10\arcmin$ largely have SF$_\infty$ and $\tau$ similar to that of stars, and are thus mostly contaminants, rather than new blazar candidates which were removed by the cut on $r < 10\arcmin$.

\begin{figure*}
\centering
\includegraphics*[width=0.90\textwidth]{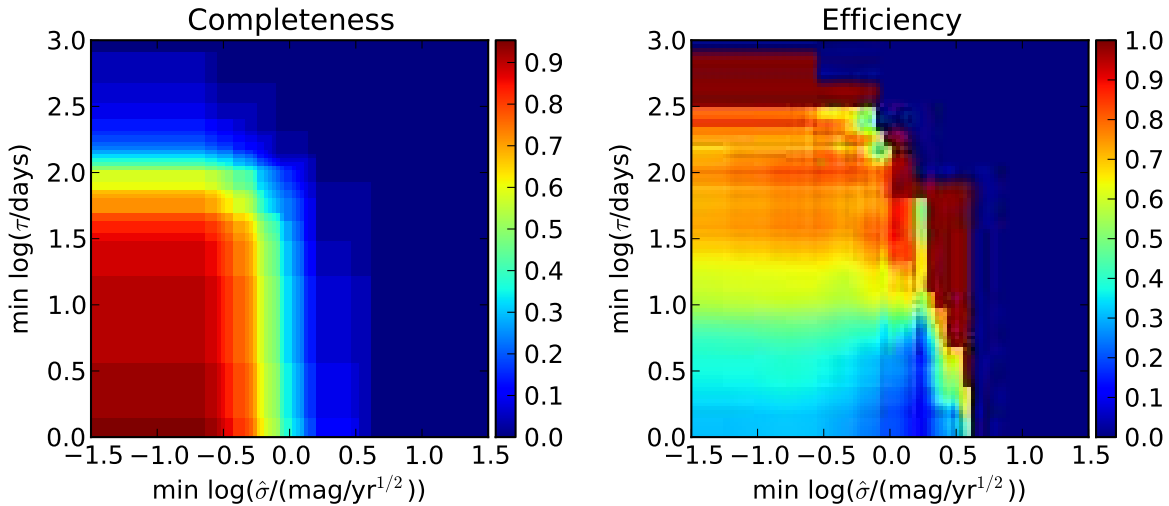}
\caption{
Completeness and efficiency of selection of confirmed \emph{Fermi} AGN in error circles of radius 
$r <10\arcmin$, as a function of selection cuts on minimum $\tau$ and $\hat{\sigma}$ . 
}
\label{fig:compeffcolor}
\end{figure*}

\begin{figure}[h!]
\centering
\includegraphics[width=0.45\textwidth]{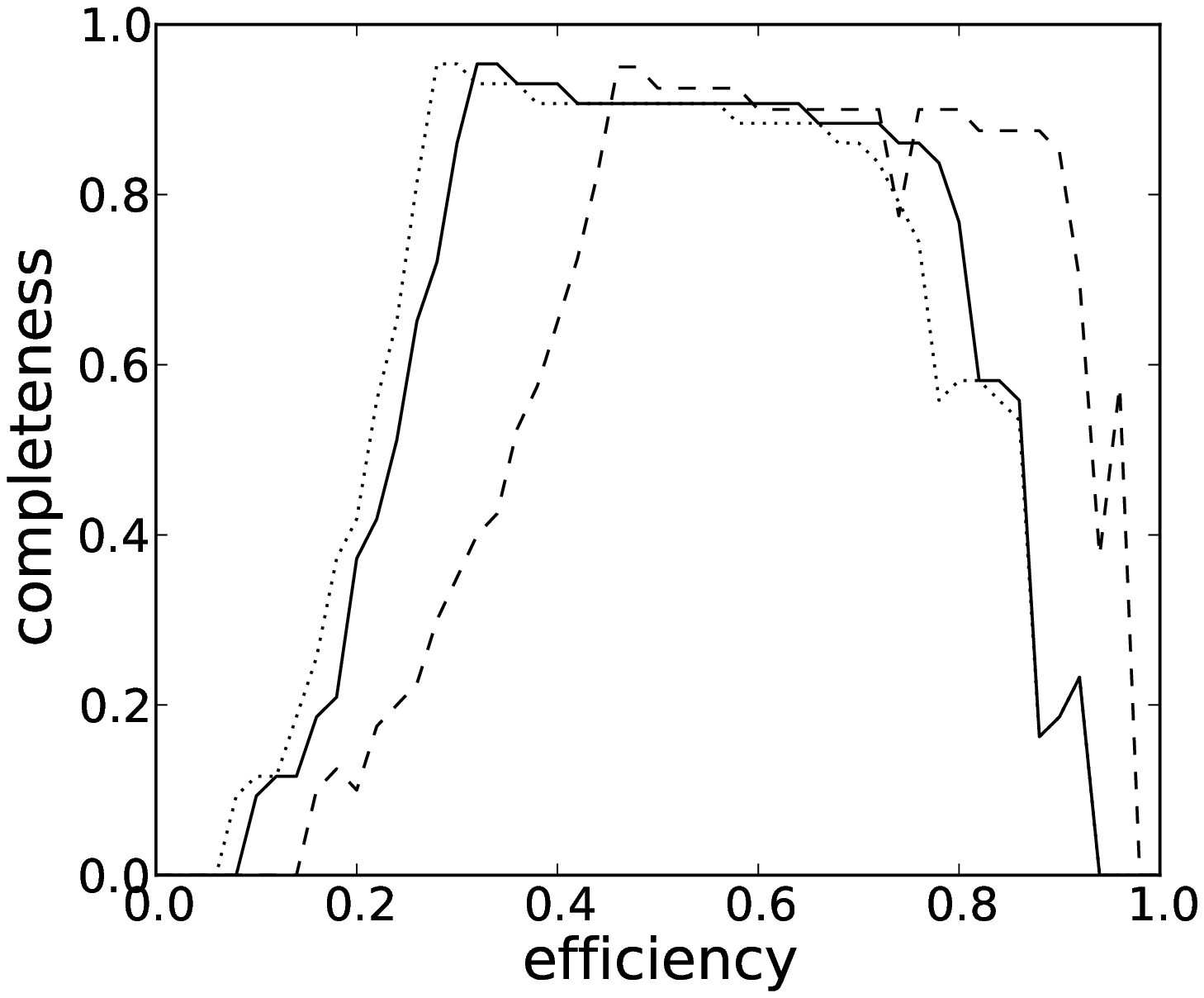}
\caption{
Maximum completeness achievable for a given efficiency of selection,  for confirmed \emph{Fermi} AGN in error circles of radius $r < 8\arcmin$ (dashed), $r < 10\arcmin$ (solid), and $r < 12\arcmin$ (dotted) as a function of efficiency.
}
\label{fig:compeffline}
\end{figure}

	We can select $\gamma$-ray AGN by placing cuts on minimum $\tau$ and 
$\hat{\sigma}$. For different log $\tau$ and 
log $\hat{\sigma}$, we estimate 
efficiency and completeness of $\gamma$-ray AGN selection. The results for a $10\arcmin$ 
cut on the error circles are shown in Figure~\ref{fig:compeffcolor}. Selection cuts at large
minimum $\tau$ and minimum $\hat{\sigma}$ yield efficiencies of 0 as there are no 
such objects. Figure~\ref{fig:compeffline}
shows the maximum completeness achievable by tweaking the selection cuts on 
minimum $\tau$ and $\hat{\sigma}$, as a function of efficiency. This shows how the 
completeness changes as a function of efficiency in $\gamma$-ray AGN selection, 
for error circles with a maximum radius of $8\arcmin$, $10\arcmin$, and $12\arcmin$. 
As expected, the efficiency decreases as the maximum size of error circles considered 
increases, at a set completeness. 

	By jointly maximizing efficiency and completeness via the quantity 
$\sqrt{E^2+C^2}$ from the curves in Figure~\ref{fig:compeffline}, we can achieve
$E$ $\ge$ 88\% and $C$ $=$ 88\% for \emph{Fermi} error circles of $r < 8\arcmin$ using cuts 
on log $\tau >1.35$ and log $\hat{\sigma} > -1.20$. For error circles of $r < 10\arcmin$, 
$E$ $\ge$ 76\% and $C$ $=$ 86\% using cuts on log $\tau >1.53$ and log $\hat{\sigma} > -0.90$,
and for error circles of $r < 12\arcmin$, $E$ $\ge$ 70\% and $C$ $=$ 86\% using cuts on 
log $\tau >1.53$ and log $\hat{\sigma} > -0.90$. This demonstration 
verifies that we are able to select (in this case, recover in double-blind fashion) 
$\gamma$-ray emitting AGN in \emph{Fermi} error circles with high completeness and efficiency 
by modeling their optical light curves as a DRW, and imposing selection cuts on the 
variability parameters. 

	Our definition for completeness is similar to the convention of many other recent 
time-domain AGN studies (e.g. MA11). The completeness we calculate here is specifically for 
the selection of variable LINEAR objects matched to \emph{Fermi} AGN in \citet{ac11} with Bayesian 
association probabilities $>$0.8, and the efficiency is calculated assuming these associations are all
correct. However, this efficiency is a lower bound, as it will 
increase if some of the contaminants we encounter in selection are
actually blazars. This may occur in our analysis if some LINEAR variables that pass all 
DRW blazar selection cuts and did not match to a AGN in the 2nd Fermi AGN catalog  
may actually be additional $\gamma$-ray AGN counterparts, not already recognized as such. 
Indeed, this is likely the case for at least one object, lying in the error circle of \emph{Fermi} 
source 2FGL J1649.6+5238.

	In the 2nd \emph{Fermi} AGN catalog, \emph{Fermi} source 2FGL J1649.6+5238 is listed as 
associated with the radio source 87GB 164812.2+524023 at a 0.0 Bayesian probability, calculated
based on the local density of sources from catalogs of likely counterparts. This radio source is 
positionally coincident with a LINEAR object at RA = 16$^h$49$^m$24$^s$.99, 
Dec = 52$^\circ$35$\arcmin$15$\arcsec$.05, 
which has DRW parameters calculated from its LINEAR light curve of 
log $\tau$ = 1.936, log $\hat{\sigma} = -0.026$, and log SF$_\infty$ =  0.942, very similar to 
that of confirmed $\gamma$-ray AGN in Figure~\ref{fig:sfinftaufermi}. This LINEAR source was 
not counted as a $\gamma$-ray AGN in our calculations of efficiency and completeness because
its Bayesian probability of association was below 0.8 in \citet{ac11}. Furthermore, this LINEAR
counterpart also did not match to any known blazar in the BL Lac object and quasar catalogs used 
in Section 2.2.

       Based on these blazar-like DRW parameters calculated from the
light curve, we suggest this LINEAR variable object as the plausible optical
counterpart to \emph{Fermi} source 2FGL J1649.6+523, as well as increased
confidence in associating 87GB 164812.2+524023 as the radio counterpart.
We note that this \emph{Fermi} source is also associated with this radio
source with a 0.82 probability using the $\log N$--$\log S$ method in
\citet{ac11}, based on observed properties of candidate radio
counterparts. The origin of this disparity lies in the differences in the
approaches used to calculate the Bayesian and $\log N$--$\log S$ (radio)
association probabilities. We opted to use the Bayesian probability $>$0.8
criterion for known $\gamma$-ray AGN, but the use of the $\log N$--$\log S$
probability also recovers this radio source as a likely $\gamma$-ray
AGN. It is not our intention to compare the two statistical approaches,
but rather to point out that the discrepancies in the results of these two
methods may lead to different results in analysis. In cases where these
two methods lead to highly discrepant results, our completely independent
optical variability-based method can provide valuable additional information.

	Finally, we note that our variability-based approach may be further applicable to 
mass identification of unassociated $\gamma$-ray sources using deeper time-domain 
optical surveys. Data from current surveys such as the PTF, with single-epoch depth 
of $r \sim 20.6$ mag, 30 epochs of observation over 8,200 deg$^2$ of sky, and $\sim$0.01 mag repeatability
can be used to identify unassociated \emph{Fermi} blazars using the method we have
presented here. Future surveys such as LSST, with single-epoch depth of $r \sim$ 24.5 
mag over 20,000 deg$^2$ of sky and $\sim10^3$ epochs per source can be used to
vastly increase the existing sample size.
		
\section{Discussion}

\subsection{Efficiency and Completeness} 
	
	 In Section 3, we noted that blazar selection using the Variable LINEAR catalog across 
the full field of the survey (as contrasted to blazar selection only within \emph{Fermi} error 
circles) was not efficient and complete. At first, this may seem to be in stark contrast to the 
success of quasar variability selection using a similar DRW modeling approach 
\citep[e.g., MA11,][]{bu11, ko10}. This is primarily a consequence of the shallow depth of the 
LINEAR survey rather than a reflection of the ultimate performance expected from our 
approach (especially for fainter blazars). At the depth of SDSS Stripe 82 (flux limit of $g\sim20.5$), 
objects with low-redshift quasar-like colors account for $\sim$63\% of all variable objects, with 
Galactic stars making up the vast majority of the remaining variable objects \citep{se07}. 
In contrast, at the shallow depth of the LINEAR survey (flux limit of $r \sim 17$), only 
$\sim$0.08\% of objects in the variable LINEAR catalog are quasars (see Section 2.3). 
As such, quasar selection by imposing 
appropriate maximum $\hat{\sigma}$ and minimum $\tau$ cuts (as done in MA11) on the 
LINEAR quasar sample in Figure~\ref{fig:sfinftauvar} 
will not yield $E$ beyond a few percent for any value of $C$ $>$ 50\% due to 
scatter in the parameters of the enormously larger population of variable stars. Blazar 
selection in LINEAR will be similarly inefficient. These issues are exacerbated by the 
large errors on the estimated DRW parameters as compared to SDSS, caused by the 
larger photometric uncertainties in LINEAR. In any case, current and 
future time-domain surveys such as PTF and LSST will alleviate both these issues 
by going many magnitudes deeper while providing $\sim$1\% photometry.
	
	As noted in Section 4, our calculated efficiencies of blazar selection 
are lower limits, since calculating the true efficiency requires correct identification of 
every object as either a blazar or contaminant. There are many contaminating variable 
LINEAR objects that fall in the \emph{Fermi} error circles we considered in our 
analysis in Section 4 that have blazar-like variability (i.e. have DRW parameters 
similar to blazars and are counted as contaminants when jointly optimizing 
completeness and efficiency for known $\gamma$-ray blazar selection). 
Some of these contaminants are certainly blazars, and may actually be the 
$\gamma$-ray source. Securing identification of these blazar-like variable objects will require 
spectroscopic follow-up. We list in Table~\ref{tab:fermicand} the positions, 
parent 2FGL error circle, and DRW parameters of 12 such new candidate LINEAR variable objects
with blazar-like optical variability that fall in \emph{Fermi} 2FGL error circles with $r <10\arcmin$. 
We also list any previously-associated counterparts of each parent 2FGL source from \citet{ac11}
with Bayesian association probabilities $>$0.8. The LINEAR blazar candidates are not positionally
coincident with these previously-associated counterparts, but are rather additional
$\gamma$-ray blazar candidates in their respective error circles. Since it is possible that some 
\emph{Fermi} error circles may contain more than one $\gamma$-ray emitting object, these LINEAR 
blazar candidates are still worthy of spectroscopic follow-up.

\begin{deluxetable*}{cccccc}
\tablecolumns{6}
\tablecaption{J2000 Positions of new candidate LINEAR blazars in \emph{Fermi} error circles with $r < 10\arcmin$ selected by  variability
\label{tab:fermicand}}
\tablehead {\colhead{RA (degrees)} & \colhead{Dec (degrees)} & \colhead{\emph{Fermi} 2FGL Source} & \colhead{log($\tau$/days)} & \colhead{log($\hat{\sigma}$/(mag yr$^{-1/2}$))} & \colhead{2nd Fermi AGN Catalog Association}}
\startdata
$128.5027$ & $42.2570$ & 2FGL J0834.3+4221 & $1.89$ & $-0.20$ &OJ 45\\
$153.3687$ & $34.4883$ & 2FGL J1013.6+3434  & $1.58$ & $-0.28$ & unassociated\\
$172.2141$ & $-5.6568$\phantom{-} & 2FGL J1129.0$-$0532 &  $2.20$ &  \phantom{$-$}$0.02$ & unassociated\\
$189.5802$ & $-20.0237$\phantom{$-$} & 2FGL J1238.1$-$1953 & $2.07$ & $-0.11$ & PMN J1238-1959\\
$219.1272$ & $23.2824$ & 2FGL J1436.9+2319 &  $1.66$ & $-0.29$ & PKS B1434+235\\
$220.4843$ & $43.9407$ & 2FGL J1442.0+4352 & $2.00$ & $-0.26$ & BZB J1442+4348\\
$232.8992$ & $57.4483$ & 2FGL J1531.0+5725 &  $2.50$ & $-0.04$ & unassociated\\
$232.9009$ & $57.3541$ & 2FGL J1531.0+5725 & $2.36$ & $-0.58$ & unassociated\\
$246.9883$ &  $32.4363$ & 2FGL J1627.8+3219 & $2.30$ & $-0.34$ & unassociated\\
$250.2259$ & $11.7870$ & 2FGL J1641.0+1141 & $1.89$ & $-0.20$ & MG1 J164058+1144\\
$252.3542$ & $52.5875$ & 2FGL J1649.6+5238 & $1.93$ & $-0.02$ & unassociated\\
$331.7412$ & $-0.4681$\phantom{-} & 2FGL J2206.6$-$0029 & $1.80$ &  \phantom{$-$}$0.27$ & PMN J2206-0031\\
\enddata
\end{deluxetable*}

	We further note that the efficiencies we have calculated in $\gamma$-ray 
blazar selection in Section 4 are conservative, as we 
have considered all variable LINEAR objects lying in all \emph{Fermi} 
error circles below a certain size. However, $\sim$10\% of sources in the 
2FGL are associated with non-AGN 
objects (e.g. pulsars, supernovae remnants, etc.), while $\sim$30\%
of sources are unassociated. We chose to include
\emph{all} these error circles in our analysis (and not just those containing known 
\emph{Fermi} blazars) in a double-blind test, as it is a more faithful demonstration
of the efficacy of this approach.

	We would like to compare the efficiency and completeness of our 
variability-based \emph{Fermi} blazar selection method to other methods.
However, a direct comparison is difficult; few studies have attempted to use 
characteristic properties of blazars to systematically select \emph{Fermi} 
blazars in double-blind fashion as we have done. Furthermore, 
the completeness and efficiency of those methods are almost 
never jointly optimized (as done in Section 4). For example, \citet{mas11}
used mid-IR color selection to recover known 
blazars in the \emph{WISE} survey, and extended their approach to find new 
\emph{Fermi} blazar candidates \citep{mas12}. By parametrizing the similarity 
of the mid-IR colors of \emph{WISE} sources to \emph{WISE}-detected \emph{Fermi} 
blazars, \citet{mas12} imposed a mid-IR color-based selection criterion that 
is able to select known \emph{Fermi} blazars with 87\% completeness over the full 
region of sky surveyed by \emph{WISE} in its first year. However,
 \citet{mas12} did not jointly calculate the completeness and efficiency of their
approach. Their mid-IR color selection method has the advantage of 
the all-sky coverage of the \emph{WISE} survey. However, mid-IR color selection 
is dependent on the degree to which non-thermal jet emission dominates the mid-IR 
emission of individual blazars. Indeed, \citet{mas11} find that FSQRs generally
 have \emph{WISE} colors closer to normal quasars than to BL Lac objects. This may be due 
 to significant thermal mid-IR emission in the SEDs of FSRQs, similar to that in normal 
 quasars \citep[e.g. from dust emission, see][]{pl12}, thus making blazar selection more difficult.

	\citet{mas12} apply their selection 
method to  \emph{WISE} sources in unidentified \emph{Fermi} error circles to
find 297 $\gamma$-ray blazar candidates in 156 out of 313 unidentified 
\emph{Fermi} error circles. However, their estimates of the efficiency of their method by 
systematically offsetting the position of each \emph{Fermi} error circle to assess 
the number of random associations gave 262 false blazar candidates (using the 
same mid-IR color selection criteria), suggesting that the efficiency of this 
approach may be low. Nevertheless, mid-IR color- and variability-based 
selection of blazars are independent and complementary methods, with
different selection biases. A combination of both approaches, using the 
full \emph{WISE} survey as well as data from 
current time-domain optical imaging surveys, may be highly 
efficient and complete.  
		
\subsection{Implications for Blazar Variability}

	Our finding that optical blazar variability is well described as a DRW process with 
distinct variability parameters may have interesting potential implications for blazar jet 
physics. The SF$_\infty$ and $\tau$ DRW parameters should both be affected 
by relativistic effects. For example, the observed blazar characteristic damping 
timescale $\tau_{blz,obs}$ (after correcting for cosmological redshift) 
should be shortened in comparison to the interval $\tau_{blz,rest}$ in the rest-frame 
of the emitting material due to the relativistic motion, with
$\tau_{blz,rest}$ =  $\delta\tau_{blz,obs}$, where $\delta$ is the kinematic 
Doppler factor. While $\tau_{blz,rest}$ 
is uncertain and difficult to measure, we can place constraints on it through
comparisons to $\tau_{qso}$, the characteristic DRW timescale of normal quasars 
(which are not affected by relativistic effects), and independent measurements 
of $\delta$. 
	
\begin{figure}
\centering
\includegraphics[width=0.47\textwidth]{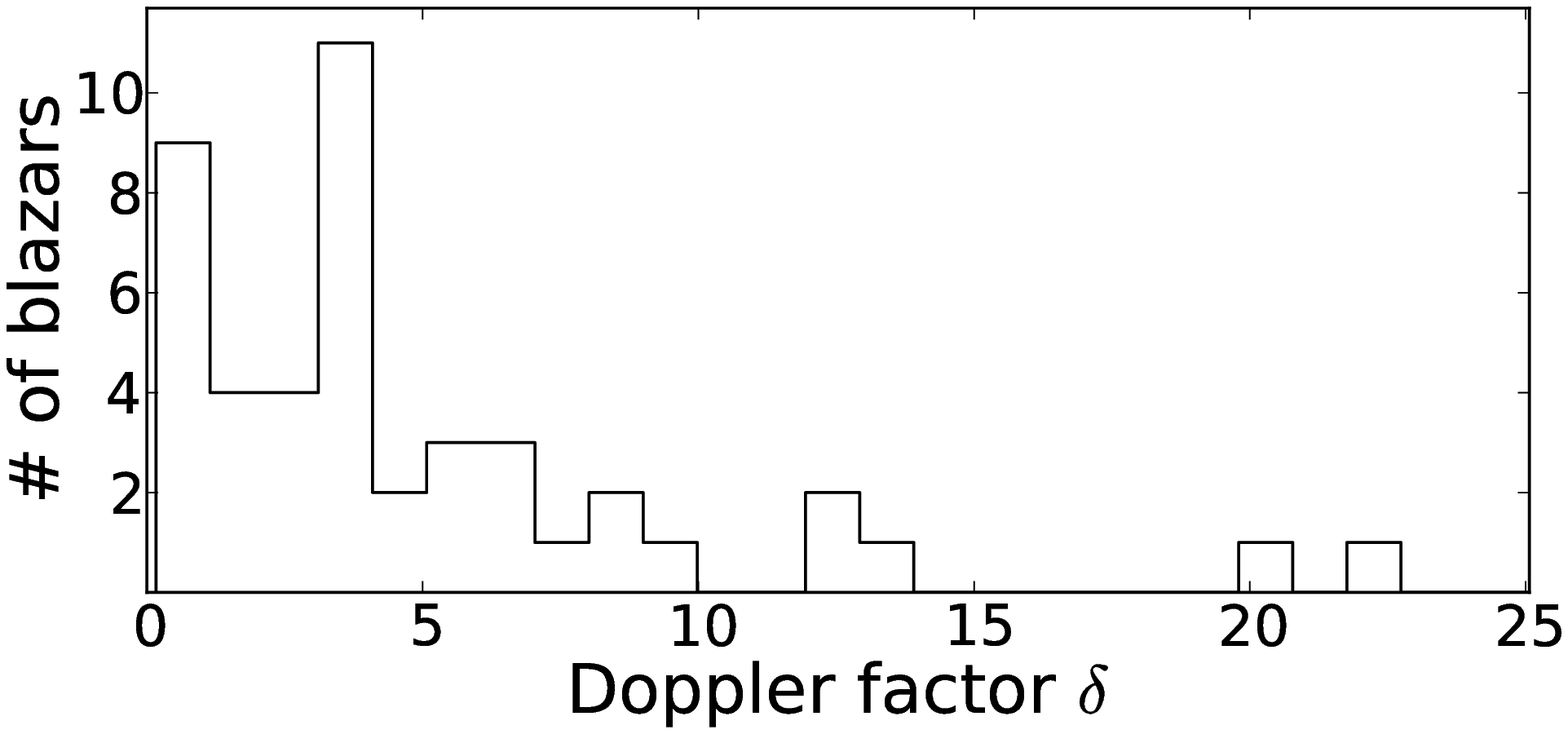}
\includegraphics[width=0.47\textwidth]{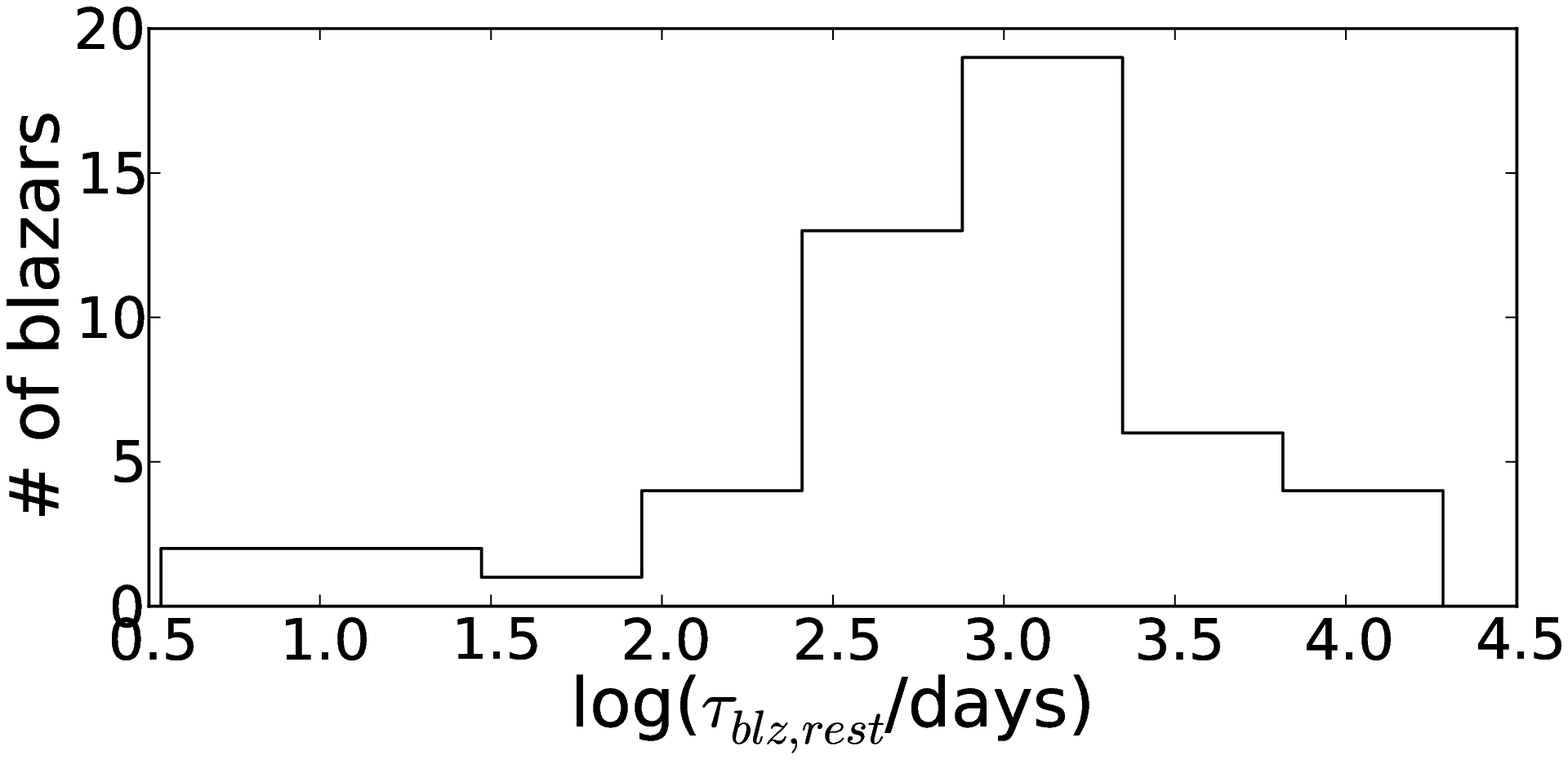}
\caption{
$Top$: Distribution of calculated kinematic Doppler factors of blazars in the Variable LINEAR catalog, assuming that the rest-frame characteristic timescale of variability for each blazar is equal to that of LINEAR normal quasars (320 days).
$Bottom$: Distribution of calculated rest-frame characteristic timescale of variability for blazars in the Variable LINEAR catalog, assuming a kinematic Doppler factor $\delta = 10$. 
}
\label{fig:doppler}
\end{figure}

	In Figure~\ref{fig:sfinftauvar}, the distributions of $\tau$ peaks 
at $\tau_{blz,obs} \sim80$ days for blazars, and $\tau_{qso} \sim 320$ days for 
normal quasars. While these distributions have large tails, their relative peaks 
provide an adequate basis for comparison. If the underlying variability in quasars
ultimately also drives blazar variability, 
then $\delta \sim \tau_{qso}/\tau_{blz,obs} \sim$ 4. 
Figure~\ref{fig:doppler} shows the distribution of this estimated 
$\delta$ for blazars in the Variable LINEAR catalog, assuming $\tau_{qso} = 320$ days. 
This is smaller than estimates of the typical blazar Doppler factor from radio observations
of \emph{Fermi}-detected blazars \citep[$\delta \sim$ 10 - 30,][]{sa10}, and may imply that the 
variability of the non-thermal jet emission in the jet 
rest-frame (e.g. due to shocks in the jet) occurs on longer timescales 
than the thermal disk variability of normal quasars.
		
	We can instead adopt the observed typical Doppler factors for blazars and 
calculate $\tau_{blz,rest}$ for each individual blazar. Ideally, we would use a measured 
value of $\delta$ for each individual blazar in this calculation, but direct
measurements of $\delta$ from the literature are sparse and inhomogeneous.
In Figure~\ref{fig:doppler} (bottom panel), we show the 
distribution of $\tau_{blz,rest}$ assuming a reasonable $\delta = 10$ for all blazars. 
The peak in Figure~\ref{fig:doppler} occurs at $\tau_{blz,rest} \sim 3$ years, which is
longer than the $\tau_{qso}\sim$ 320 days we measure for LINEAR normal quasars.
The discrepancy between $\delta$ for \emph{Fermi} blazars implied by radio observations 
and $\delta$ calculated assuming intrinsic variability similar to normal quasars suggests 
different stochastic mechanisms driving the variability in the disk and the jet.
While a detailed physical interpretation of this is currently unclear, 
our long-term optical variability results may provide additional constraints on models 
of blazar jets. 

\subsection{Short Timescale Variability of Quasars and Blazars}

	Observations of intraday variability (also referred to as `microvariabilty') of 
normal quasars have shown a puzzling diversity of properties. Not all quasars 
exhibit this phenomenon, but some show variability on the order of 
$\sim$0.01 mag on $\sim$1 day timescales \citep{go03, sta04, sta05, gu05, mu11}. 
The cause of quasar variability on such short timescales is an unresolved 
problem, although the presence of strong intraday 
variability is correlated with radio-loudness, and it has long 
been suspected that a weak jet component may 
be both the source of radio emission and the cause of the stronger variability. 
\citet{ke09} showed that the DRW model successfully predicts amplitudes of variability of 
$\lesssim$0.02 mag over $\sim$8 hours for quasars, consistent with observations from quasar
monitoring. This successful prediction of the short-timescale variability  is
noteworthy since the sampling intervals of the quasar light curves used in 
\citet{ke09} are all $\geq$2 days, and thus do not actually sample such intraday 
timescales.

	From the log $\hat{\sigma}$ distributions for normal quasars and blazars 
in Figure~\ref{fig:sigmahist}, we can calculate the standard deviation of the expected 
variability on $\sim$1 day timescales, approximated as $\hat{\sigma}$$\sqrt{\Delta t}$. 
For quasars, the peak of the distribution occurs at $\sim$0.2 mag yr$^{-1/2}$, thus 
predicting intraday variability $\sim$0.01 mag on 1 day timescales. For blazars, the 
peak occurs at $\sim$1 mag yr$^{-1/2}$, thus predicting variability of $\sim$0.05 mag 
on 1 day timescales in the observed frame. Previous studies of blazar microvariability 
have often focused on the most variable blazars, and have reported  
night-to-night variations as high as 1.0 mag \citep[e.g.][]{ca91, go00}. The intraday
variability we have calculated for LINEAR blazars is from an untargeted (i.e. less biased) 
survey, and is likely more representative of the intraday variability of blazars
as a whole. 

	Despite the large photometric uncertainties in the LINEAR 
survey, there are a significant number of normal quasars and blazars in Figure~\ref{fig:sigmahist} 
which do have extreme levels of short-timescale variability, above the photometric 
uncertainty. Nearly all LINEAR objects have photometric uncertainties below 0.1 mag
(and reaching as low as $\leq$0.04) at $<$17 mag (see Figure~\ref{fig:photoerrors}). 
Conservatively assuming that variability on 1 day timescales needs to be $>$0.1 mag 
in order to be detectable above the photometric uncertainty implies a
log $\hat{\sigma}$ $>$ 0.28. In Figure~\ref{fig:sigmahist}, 18 of 119 normal quasars and 
10 of 51 blazars in our sample are variable above this criterion (and both these fractions would 
increase if we more carefully considered the 
photometric uncertainty limit of each individual object). Thus for these objects, the intraday 
variability is above the LINEAR photometric uncertainty and the $>$0.1 mag
intraday variability is significantly above most previous observations of normal quasars. 
The source of this extreme variability for a subset of quasars is unknown, and warrants 
additional investigation. 

\section{Conclusions}

	We have used light curves from the LINEAR optical imaging survey to study the 
variability of blazars, and to compare their variability to that of normal quasars. In contrast 
to many earlier studies of blazar variability, this study is the first to use a reasonably large, 
homogeneous sample of individual light curves from a wide-field 
survey, where the distribution of various variability properties for blazars and quasars 
can be calculated and compared. 	
	
	In a time-domain survey, blazars are among the most violently variable objects 
detected, and are generally more variable than normal quasars. We estimate that 
$\sim$36\% of blazars but only $\sim$11\% of quasars are variable at the extreme $>$0.15 mag intrinsic rms 
variability level. We show that, like quasars, blazar light curves are well-fit by the DRW 
model for variability, but the blazars lie in distinct variability parameter-space with higher $\tau$ 
than stars and higher $\hat{\sigma}$ than normal quasars. This suggests that blazars 
can be selected with high efficiency $E$ and completeness $C$ by 
imposing minimum selection cuts on $\tau$ and $\hat{\sigma}$. 
	
	Due to the overwhelming numbers of variable stars in the bright magnitude regime 
probed by LINEAR, it is difficult to select blazars (and quasars) at high efficiency and 
completeness in the full LINEAR sample using optical variability methods alone. 
We instead examine a more focused test to select $\gamma$-ray emitting blazars in \emph{Fermi} 
error ellipses from the 2FGL catalog. We calculate DRW parameters for light curves 
of all LINEAR variable objects objects within \emph{Fermi} error ellipses and place 
minimum selection cuts on $\tau$ and $\hat{\sigma}$. Using this approach, we are able to recover the 
corresponding $\gamma$-ray emitting AGN counterparts in the 2nd \emph{Fermi} 
AGN catalog with $E$ $\ge$ 88\% and $C$ $=$ 88\% for 95\% confidence error ellipses 
with semimajor axis $r < 8\arcmin$, and $E$ $\ge$ 70\% and $C$ $=$ 86\%  for $r < 12\arcmin$. 
	
	Our $\gamma$-ray blazar selection has uncovered a variable LINEAR object, 
coincident with radio source 87GB 164812.2+524023, in the error ellipse of 
\emph{Fermi} source 2FGL J1649.6+5238. This object was not associated with the 
\emph{Fermi} source in the 2nd \emph{Fermi} AGN catalog using Bayesian 
probabilities. Our analysis shows this object has optical variability properties
consistent with $\gamma$-ray emitting blazars and is likely to be the $\gamma$-ray
source. We find a total of 12 objects with LINEAR variability parameters similar to 
blazars lying in \emph{Fermi} error circles with $r < 10\arcmin$ (see Table 1). 
Confirmation of these variability-selected blazar candidates will require 
spectroscopic follow-up.

 	Our results suggest that the variability of the non-thermal jet emission in 
blazars is stochastic in nature, with higher amplitudes and shorter observed timescales 
in comparison to the variability of normal quasars, likely due to the effects of relativistic 
beaming. Assuming a reasonable Doppler factor of 10, we estimate that blazars are 
characteristically variable on timescales of $\sim$3 years in the rest-frame of the 
emitting material, longer than the $\sim$320 days characteristic disk flux timescale for quasars.
This suggests that different physical mechanisms dominate the observed variability from blazars 
and quasars, likely connected to the jet and to the disk, respectively.
The fitted parameters imply that blazars have 
a typical intraday variability amplitude of $\sim$0.05 mag, 
compared to $\sim$0.01 mag for normal quasars. Furthermore, 
there is a significant fraction ($\sim$15\%) of normal 
quasars which exhibit large intraday variability of $>$0.1 mag, detectable above the 
photometric uncertainty. The source of the extreme variability of these quasars is unclear. 

	We argue that variability-based blazar selection is likely to be 
highly efficient and complete in deeper optical time-domain imaging surveys, and that
the variability-based blazar selection method presented in this paper is capable of 
greatly increasing the number of known blazars. Our approach, based on long-term 
photometric variability characteristics of a large sample of individual blazar light curves, 
may also bring a new perspective on accretion disk and jet physics. Finally, this work 
typifies the fruitful ancillary science made possible by combining data from surveys
as diverse in all respects as LINEAR, SDSS, and \emph{Fermi}.

\acknowledgments
	The authors wish to thank Eric Agol (U. of Washington) and Ying Zu (Ohio State U.) for helpful 
discussions regarding AGN variability. {\v Z}.I. and C.L.M. acknowledge support by NSF grant 
AST-0807500 to the University of Washington, NSF grant AST-0551161 to LSST 
for design and development activity, and Croatian National Science Foundation 
grant O-1548-2009. R.M.P. acknowledges support from a Netherlands 
Organization for Scientific Research (NWO) Vidi Fellowship. C.S.K. is supported by NSF 
grant AST-1009756. B.S. acknowledges NSF grant AST-0908139 awarded to Judy Cohen for 
partial support.

	The LINEAR program is sponsored by the National Aeronautics and Space 
Administration (NRA No. NNH09ZDA001N, 09-NEOO09-0010) and the United States 
Air Force under Air Force Contract FA8721-05-C-0002. Opinions, interpretations, conclusions, 
and recommendations are those of the authors and are not necessarily endorsed by the 
United States Government.

\bibliography{bibref}
\bibliographystyle{apj}

\end{document}